\documentstyle[multicol,epsf,aps,prl,twoside]{revtex} 
\begin{document}

\title{Size effect on magnetism of Fe thin films in Fe/Ir superlattices}

\author{S. Andrieu$^{1}$, C. Chatelain$^{2}$,  M. Lemine$^{1}$, 
	B. Berche$^1$, and Ph. Bauer$^{3}$}
\address{$^1$Laboratoire de Physique des Mat\'eriaux,
	Universit\'e Henri Poincar\'e, Nancy 1, BP 239,
	F-54506  Vand\oe uvre, France\\ 
	$^2$Institut f\"ur Theoretische Physik, Universit\"at Leipzig,
	D-04109 Leipzig, Germany\\
	$^3$CREST, Universit\'e de Franche-Comt\'e,
	BP 71427 25211 Montb\'eliard Cedex, France\\
}
       
\date{\today}
\maketitle

\begin{abstract}
In ferromagnetic thin films, the Curie temperature variation with the 
thickness is always considered as continuous when the thickness is varied 
from $n$ to $n+1$ atomic planes. We show that it is 
not the case 
for Fe in Fe/Ir superlattices. For an integer number of atomic planes, a 
unique magnetic transition is observed by susceptibility measurements, 
whereas two 
magnetic transitions are observed for fractional numbers of planes. This 
behavior 
is attributed to successive transitions of areas with $n$ 
and $n+1$ atomic planes, for which the $T_c$'s are not the same.  
Indeed, the magnetic correlation length is presumably shorter than 
the average size of the terraces. Monte carlo 
simulations are performed to support this explanation. 
\end{abstract} 

\begin{multicols}{2}
\narrowtext
 
Motivated 
by both the richness of physical phenomena and the potentiel applications in 
magnetoresistive heads, sensors, and more recently spin electronics,
there has been considerable progress  
in the area of thin magnetic films during the last fifteen years. 
The control of 
the reduced dimension of the devices, including thin films, multilayers or 
magnetic dots, became crucial. 
The progress realized in deposition techniques like 
sputtering or molecular beam epitaxy (MBE) made possible the deposition of 
thin films
of a thickness of several atomic planes with a very good
accuracy and reproductibility. Lithography on the other hand
is used to reduce the lateral dimension. This accurate control 
of the sizes is a necessary condition for preparing model 
systems in order to investigate, for 
instance, the magnetic coupling between two magnetic layers via a spacer, 
the coherent rotation of a single magnetic domain with an applied magnetic 
field,  the spin 
dependent tunnel current via an oxide,\dots For instance, the 
well-characterized thin films prepared in UHV conditions can be considered as 
model systems for studying the variation of the magnetic ordering 
temperature with the 
number of deposited atomic planes. This was done for a number of systems -Co, 
Fe, Ni, FeNi and CoNi on Cu~\cite{SchneiderEtAl90,BallentineEtAl90},
 Gd on (0001) Y~\cite{GajdzikEtAl98}, Fe on (001) Ir~\cite{HenkelEtAl98} 
and Pd~\cite{ChoiEtAl99}. 
The experimental results were often compared to scaling laws~\cite{BallentineEtAl90,LiEtAl92,GajdzikEtAl98,HenkelEtAl98}, which
predict 
a universal dependence of the reduced Curie temperature with the number of 
atomic planes. Power laws are obtained, with universal critical exponents, 
which depend on the type of magnetic interactions (the so-called universality 
class). In all previous experimental studies, 
the Curie temperature was assumed and observed to continuously vary with 
the quantity of deposited material, even for thicknesses corresponding to a 
fraction of atomic plane. This is actually true keeping in mind that the 
magnetic correlation 
length $\xi$ diverges at the magnetic transition temperature in 
the bulk. However, this argument becomes questionable for systems with reduced 
dimensions. For a fractional number of deposited atomic planes between $n$ and 
$n+1$, if the growth occurs layer by layer, the film is made of  
areas of $n$ atomic planes while other areas have $n+1$ atomic 
planes, both areas 
having separately different ordering temperatures, provided that their
lateral extent exceeds the correlation length. If $\xi$ is lower than the 
typical size of these areas, 
it is thus no more justified to assume a unique transition. 
In this Letter, the observation of two 
magnetic transitions in a continuous iron film is reported. This effect is 
actually explained by reduced magnetic coherence length as shown by Monte 
Carlo simulations.

\begin{figure}
	\vspace{-0mm}
	\epsfysize=7cm
	\begin{center}
	\mbox{\epsfbox{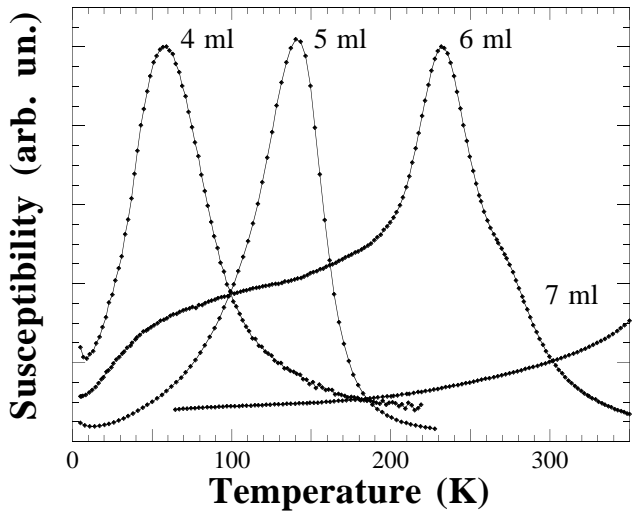}}
	\end{center}
	\vskip 0cm
	\caption{ac susceptibility peaks observed for Fe/Ir multilayers 
	of 20 periods with Ir spacer thickness of 15~\AA, and
	integer numbers of 
	Fe atomic planes.}
	\label{Fig2}  
\end{figure}

The preparation and characterization of the Fe/Ir(100) superlattices (SLs) 
were already reported in previous papers~\cite{HenkelEtAl98,AndrieuEtAl95,AndrieuEtAl95bis,Lemine99}. We thus briefly 
describe the main properties of iron in these superlattices. The epitaxial 
growth of Fe on (001) Ir is performed by MBE at 400~K. The Fe growth mode on 
Ir(100) is Stransky-Krastanov. Indeed, as layer by layer growth 
takes place up to 5-6 atomic planes, three-dimensional (3D) growth starts 
above this thickness~\cite{AndrieuEtAl95}. 
Using high-resolution transmission microscopy (HRTEM), 
we have shown that the SLs constituted of Fe layers up to 6 atomic planes 
thick exhibit flat interfaces from the first
to the last period, whereas interfacial roughness is observed for larger Fe
thicknesses, when a 3D growth takes 
place~\cite{AndrieuEtAl95,AndrieuEtAl95bis}. Moreover, HRTEM and
M\"ossbauer spectroscopy allow us to control that the Fe structure and
magnetic properties are the same for all the Fe layers when no misift
dislocations take place, that is before 3D 
growth~\cite{AndrieuEtAl95bis,Lemine99}.
The structural 
analyses show that Iron first grows in a body centered tetragonal (bct) 
structure up to 4 atomic 
planes. Above this thickness, the additional Fe planes are observed to grow 
in the bcc structure on top of the bct 
structure~\cite{AndrieuEtAl95bis}. The control 
of the thickness is achieved using the RHEED intensity oscillations 
during the 2D growth. The accuracy on the thickness determination is 
estimated to be better than 0.1 monolayer (ml). The Fe magnetic properties 
were investigated  using several techniques. First, a Fe magnetic ordering is 
observed for Fe thicknesses above 2~ml, and Ir thicknesses above 3~ ml, as 
shown 
by SQUID and M\"ossbauer 
measurements~\cite{AndrieuEtAl95bis,Lemine99}. The Curie 
temperature  
varies from 15~K for 3~ml up to 215~K for 
6~ml~\cite{HenkelEtAl98}. The magnetization is in-plane and a 2D spin wave 
behavior is observed for Fe thicknesses up to 6~ml~\cite{Lemine99}. 
Low angle neutron 
scattering shows that the Fe layers are not coupled via Ir 
for Fe thicknesses up to 6 atomic 
planes (even down to 4~K). This is an important point, since the order 
temperature might be affected by this coupling~\cite{BovensiepenEtAl98bis}. 
To summarize, the Fe/Ir SLs may be considered as a set of flat 
and continuous Fe layers uncoupled with each other. The present work 
completes this magnetic study, 
using ac susceptibility ($\chi$) experiments. In order to get a detectable 
signal, we work on Fe/Ir SLs with 20 periods with Fe 
thicknesses varying from 4 to 7 planes. The Ir thickness was fixed to 15~\AA. 
The Fe thickness was first accurately determined by fitting the RHEED 
oscillations. The $\chi$
 experiments were 
first performed on Fe/Ir SLs grown at 400~K with an integer number of Fe 
atomic planes from 4 to 7, but we also investigated superlattices 
grown at room temperature in order to check the roughness influence.
In the former case, well defined transition peaks are 
observed (Fig.~\ref{Fig2}), and the corresponding Curie temperatures are 
in total agreement with M\"ossbauer experiments~\cite{HenkelEtAl98}. 
Consequently, for an integer 
number of Fe atomic planes, only one magnetic transition is observed. 
However, this behavior does not persist for Fe layers consisting of a 
fraction of atomic planes. In Fig.~\ref{Fig3} are shown the $\chi$ spectra 
obtained on 
a set of SLs where the Fe thickness is varied from 4 to 5~ml by steps of 
0.2~ml. Two peaks are now present, 
which are approximately located at the transition 
temperatures observed for 4 and 5~ml,
respectively. 
We observe the same type of behavior for Fe thicknesses between 5 and 6 planes.
These results may clearly not be explained by a continuous variation 
of the Curie temperature with the Fe thickness. 
Each peak in the $\chi$ spectra arises from a 
part of the SL with exactly 4 or 5 
atomic planes. This could not be explained by some thickness variation on the 
substrate surface, since in our apparatus the thickness 
homogeneity is 2\% on a 2 inch sample. 
A thickness dispersion from one Fe layer to the other in 
the SL could also not be at the origin of these peaks since the incident Fe 
flux dispersion during the preparation of a SL is better than 5\%. We could 
also imagine that the two magnetic transitions come from the two crystalline 
bct and bcc phases present above 4~ml. However, this assumption is again not 
satisfactory. Indeed, for exactly 5~ml of Fe, both bct and bcc phases coexist, 
but only one transition is observed. This means that both phases have the 
same $T_c$ if the interface between them is infinite. 
This is confirmed by M\"ossbauer spectroscopy, where two different
magnetic phases are actually observed, but with the same hyperfine
magnetic field~\cite{Lemine99}.
This behavior is also 
observed on the Co/Ni system for instance~\cite{AspelmeierEtAl95}. This 
behavior can finally be 
explained by the morphology of a Fe growing layer. As the 
growth is layer by layer, a 4.5~ml thick Fe layer is constituted of some 
areas with 4 planes and other areas with 5 planes. The magnetic 
transition observed at 70~K can be attributed to the areas with 4~ml and the 
other transition at 140~K to the areas with 5~ml. We thus implicitly consider 
that both areas have their proper Curie temperature despite the fact that 
they are in contact. This is possible if the magnetic coherence length  
is smaller than the typical size of the terraces. 
Consequently, decreasing the distance between the $2D$ islands in the last
atomic plane up to the magnetic coherence 
length should lead to the occurrence of only one broader peak. 
\vbox{\begin{figure}
	\vskip-0mm
	\epsfysize=4.2cm
	\mbox{\epsfbox{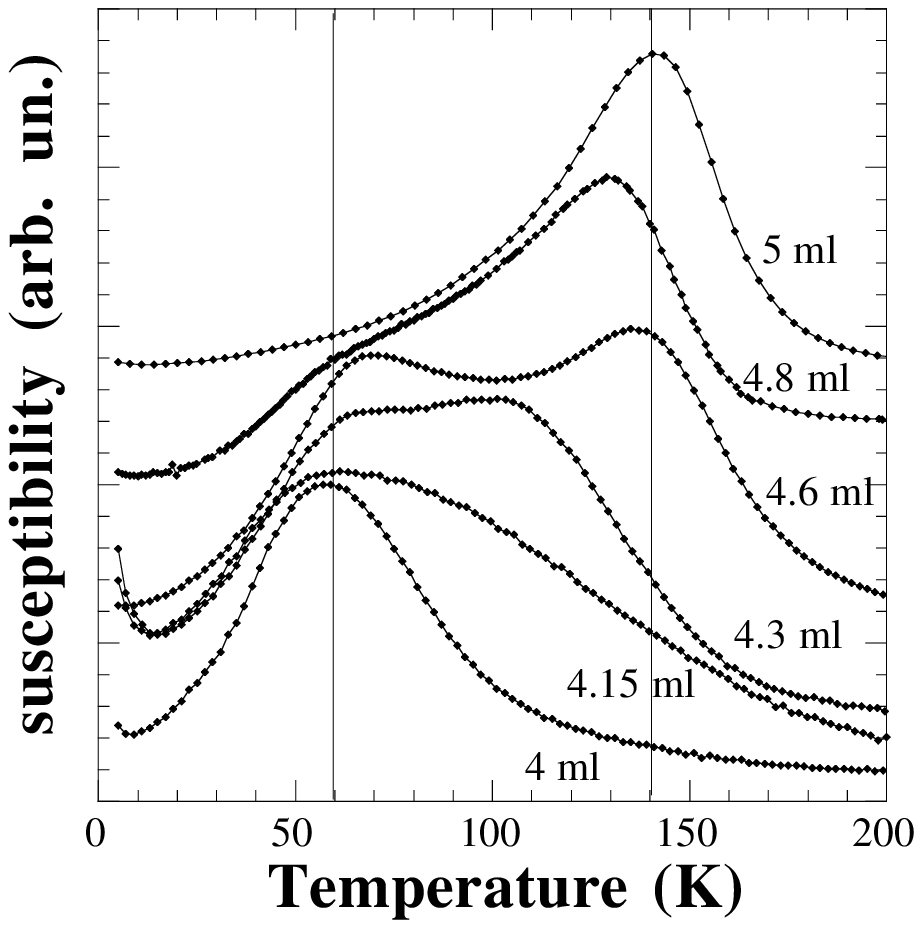}}\hfill\
	\vskip-4.2cm\epsfysize=4.2cm
	\ \hfill\mbox{\epsfbox{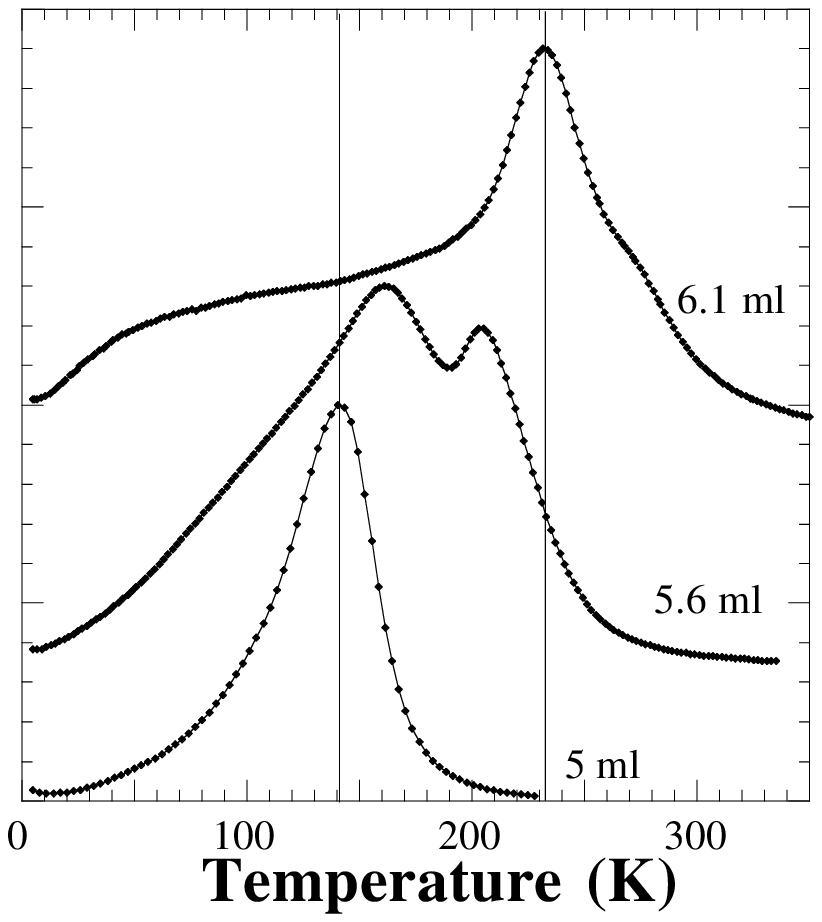}}
	\vskip 0mm
	\caption{ac susceptibility peaks observed for a set of SL with Fe 
	thicknesses varying from 4 to 5~ml (left) and from 5 to 6~ml (right).}
	\label{Fig3}  
\end{figure}
}
\begin{figure}[ht]
	\vspace{-0mm}
	\epsfysize=7cm
	\begin{center}
	\mbox{\epsfbox{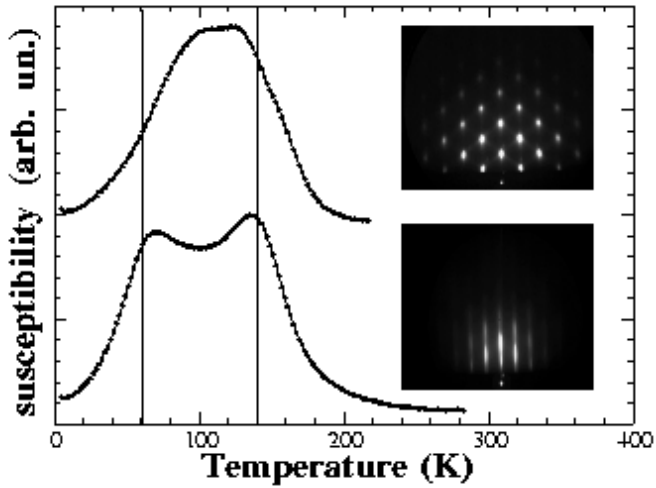}}
	\end{center}\vskip 0cm
	\caption{Influence of the surface roughness as shown by the RHEED 
	patterns (shown in insert) on the ac susceptibility measurements for 
	a SL with 4.6~ml thick Fe layers grown at 300~K (top) and 400~K 
	(bottom).}
	\label{Fig5}  
\end{figure}
The decrease of the 
islands separation is possible by decreasing the epitaxial temperature 
since the dynamical roughness is thus increased. A SL with 4.6~ml per Fe 
layer was thus grown at room temperature. As shown in Fig.~\ref{Fig5}, the 
RHEED patterns obtained at the end of the SL growth are actually typical of a 
rough surface. The $\chi$ 
analysis thus shows that, if two peaks are 
observed for the sample grown at 400~K, 
only one peak which spreads out is observed for the sample 
grown at RT, the transition 
temperature being 
thus located between the Curie temperatures of 4 and 5~ml samples. 

In the following, we report extensive Monte Carlo (MC) simulations which 
support the
previous interpretation of the double-peak structure of the magnetic
susceptibility as the signature of the presence of wide terraces in 
the magnetic
layers. These simulations do not intend to reveal the details of the structure
of the samples, but are considered as an illustration of the finite-size 
mechanism which leads to several peaks in the susceptibility signal.
For that reason, we have chosen the simplest model of magnet displaying
a ferromagnetic-paramagnetic phase transition, the Ising model. 
We
studied qualitatively the effect of a finite lateral length as well as of a
finite transverse size of the sample on the susceptibility behavior.
These effects are well described in the context of critical phenomena by the
standard finite-size scaling theory~\cite{barber83}. 
Since the Fe layers are magnetically decoupled, and because there is no need 
of any amplification of the signal in the MC simulations, it is not necessary 
to pile up layers.
We thus considered in the model a layer constituted of two parts with 
different numbers of atomic planes, let say two regions with
$n$ and $n'$ monolayers of magnetic material, which models the 
juxtaposition of wide terraces having different thicknesses. 
The boundary conditions are periodic in the plane of the sample and free in 
the transverse direction. The lattice is a $3D$ simple cubic lattice,
with Ising variables
$\sigma_{i}=0,1$ located at the sites of the lattice and interacting
via constant nearest neighbor ferromagnetic 
exchange interactions $J$. In the model, two different magnetic phases with
different coupling strengths and different cristallographic structures should
a priori be used, but since the M\"ossbauer hyperfine field is the same,
we keep the same $J$ and same structure for both phases in order to get
the same $T_c$'s.
The Hamiltonian
is written
${\cal H}=-J\sum_{(i,j)}\delta_{\sigma_{i},\sigma_{j}}.$
This model is the Potts version of an Ising model. 
In the following, we introduce the  variable
$K=J/k_BT$.
The computation of the thermodynamic properties is performed through 
MC simulations. 
Since we are interested in a second-order phase transition, the resort
to cluster update algorithms is more 
convenient, since 
it is less affected
by the critical slowing down than the conventional Metropolis algorithm. 
In order to avoid
time consuming computations at many different temperatures, we also used
a histogram reweighting technique.

For the sake of clarity, the main results
of the finite-size scaling theory are recalled
for consistency.
We first consider a simple system of shape $n\times L\times L$.
For an ideal system of finite thickness $n$ and infinite lateral extent
(thermodynamic limit $L\to \infty$ in the plane), 
the susceptibility exhibits the usual $2D$
critical behavior with respect to the
deviation $t=|K-K_c(n)|$ from the critical coupling strength $K_c(n)$, 
$\chi_\infty(K)=t^{-\gamma}$. 
The critical behavior is the same as for a single 
monolayer, but the critical coupling
strength $K_c(n)$ is reduced (or its inverse, the critical temperature,
is increased). 
In the simulations, the systems are of finite size in the three directions.
For a fixed  thickness $n$ and finite lateral extent $L$, the 
singularity of the susceptibility
expands as $L$ increases, but it remains finite at any 
temperature. This phenomenon is usually described by a 
finite-size scaling ansatz
$\chi_L(K)=t^{-\gamma}f(L/t^{-\nu})$
where $t^{-\nu}$ is
the power law behavior of the correlation length of the infinite
system and $f(x)$ is a universal scaling function.
The susceptibility reaches its maximum for some value of the scaled 
variable $x^*=Lt_{max}^\nu$. 
The condition of a non singular susceptibility for a finite sample
also implies a power law for the scaling function 
$f(x)\sim x^{\gamma/\nu}$ in the region of the finite-size effects 
$t^{-\nu}\gg L$  in order to cancel the temperature dependence.
The maximum of the susceptibility thus increases as power
of the size of the system $\chi_{max}\sim L^{\gamma/\nu}$, and this
mechanism is responsible for the absence of real divergence of $\chi$,
as observed in experimental systems where defects are always present
which stop the possible divergence of correlations.  
On the other hand the scaling function simply reaches a constant value far
from criticality when
$\xi$ is still very small compared to the typical size $L$, in a region
where $\chi$ cannot be distinguished from the susceptibility of an
infinite system.
If $L\to\infty$, the behavior is purely $2D$, but
the thickness $n$ has also to be taken into account as a scaling field
which induces a crossover towards a $3D$ behavior
when $n\gg 1$. 
Close to 
$K_c(n)$, a pure $2D$ critical singularity is thus observed
$ \chi_\infty(K,n)\sim {\rm const.}\times t^{-\gamma}$
while a dimensional crossover is obtained when $n\gg t_{3D}^{-\nu_{3D}}$ 
($t_{3D}$ is now the deviation from the bulk critical coupling,
$t_{3D}=|K-K_c(\infty)|$).
The critical coupling $K_c(n)$ for example evolves
towards the bulk critical value 
$K_c(\infty)$ according to a power law
which involves the $3D$ correlation length exponent $\nu_{3D}$.

For a sample with a step between
two terraces of $n$ and $n'=n-1$ layers and a covering rate
which varies between 0\% 
and 100\%, 
one can observe a  susceptibility with two peaks whose 
amplitudes strongly depend on the coverage and whose positions are slightly
shifted from the positions corresponding to 
pure samples at $n$ or $n'$
monolayers. 
This is shown in Fig.~\ref{fig3} for $n=5$ 
for several covering rates at $L=512$. 
The two peaks correspond to {\bf two} transitions in the thermodynamic limit
(this limit is of course not reached in the real samples where the terraces
always have a finite lateral extent).
Starting from the high temperature phase where the system is
paramagnetic, the first peak corresponds to the ordering of the thicker part of
the sample, at a temperature where the thin part is still disordered. It is
a true transition in the limit $L\to\infty$ at a given coverage. When the
temperature decreases again, the remaining part of the sample becomes
ferromagnetic and gives rise to a second transition.
The two transitions are not exactly the same: While the high temperature
peak corresponds to an ordinary transition, the low temperature transition
occurs in the presence of a ferromagnetic order in the thicker regions of
the sample. For the spins at the interface, in the vicinity of the
step, this is analogous to an  extraordinary transition 
when a surface orders while the bulk is already ferromagnetic~\cite{binder83}. 

\vbox{
\begin{figure}[ht]
\vspace{-0mm}
\epsfxsize=7cm
\noindent\hfill\mbox{\epsfbox{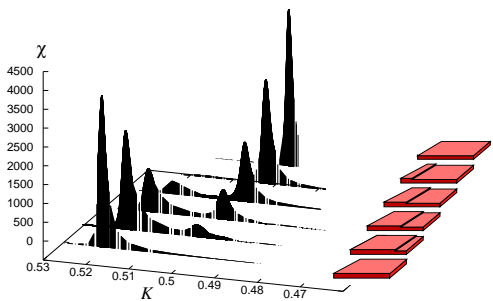}}\ \hfill
\vskip 0.1mm
\caption{Evolution of $\chi$  as the covering rate
is varied between 4 and 5~ml, 
0 \%, 20\%, 40 \%, 60 \%, 80 \%, and 100 \% from top to bottom. 
In the intermediate region, $\chi$  displays
two peaks located at 
the ``pure'' systems temperatures.
}
\label{fig3}\end{figure}
}
The scenario proposed here is confirmed by the value of the
correlation length. 
Outside criticality, the connected correlation function (after substraction
of the square of the order parameter in the ferromagnetic phase)
exhibits the usual exponential 
decay 
	$G_\sigma(r)\sim e^{-r/\xi}$.
We can use this expression to 
have an evaluation of $\xi$ at different temperatures.
For that purpose, we computed $G_\sigma(r)$
along the direction parallel to the step in the plane. 
An exponential fit thus 
leads to the 
corresponding value of $\xi$.
Starting from the paramagnetic phase, the correlation length 
remains quite small and is not strongly affected by the presence of the step.
As $K_c(n)$ is approached,
$\xi$ increases strongly (up to $12$ in lattice spacing units) 
in the corresponding region, but 
remains small ($<4$)) in the thin part of the system. 
We can thus understand that this
first peak in the susceptibility is linked to the appearence of ordering in the
thick regions of the sample, while the thin part remains paramagnetic.
When $K$ is increased again, a second singularity develops in $\chi$, 
and is associated to the ordering of this thin part where 
$\xi\simeq 6$. These observations are confirmed by
the order parameter profiles.

To summarize, we have observed that the Curie temperature of Fe in Fe/Ir SL 
does not vary continuously with the Fe quantity in the Fe layers. Between 4 
and 5 atomic planes, the Fe layer is a mixing of areas with 4~ml and areas 
with 5~ml, which leads to two magnetic transitions 
corresponding to the $T_c$'s 
of  4 and 5~ml. This behavior is explained by assuming that the lateral 
magnetic coherence length is smaller than the distance between 5~ml thick 
areas. This assumption is supported by experiments performed on SLs with 
rough interfaces. In that case, only one magnetic transition is observed. 
The MC simulations support this explanation.  Indeed, by varying 
the distance between n+1 atomic planes thick areas, the film can exhibit one 
or two Curie temperatures. This observation is in fact not surprising, but
requires high quality samples (absence or interface roughness) 
and what is more surprising is indeed that
such an observation was never reported in the literature (to the
authors knowledge).




\def\paper#1#2#3#4#5{#1, #3 {\bf #4}, \rm #5 (#2).}
\def\papers#1#2#3#4#5{#1, #3 {\bf #4}, \rm #5 (#2)}

\end{multicols}
\end{document}